\documentstyle[epsf,aps]{revtex}
\def\npb{{ \sl Nucl. Phys. }}
\def\prl{{ \sl Phys. Rev. Lett. }}
\def\plb{{ \sl Phys. Lett. }}
\def\cum#1{\langle\langle#1\rangle\rangle}

\def\half{{1\over2}}
\def\c#1{{\cal{#1}}}
\def\bfr#1{{#1}_{\bf r}}
\def\eqnn#1{(\ref{#1})}
\def\figno#1{Fig.~\ref{fig:#1}}
\def\ave#1{\left\langle#1\right\rangle}
\draft
\begin{document}
\title{
  Violations of local equilibrium and linear response\\
  in classical lattice theories}
\author{  
  Kenichiro Aoki\footnote{E--mail: {\tt ken@phys-h.keio.ac.jp}}
  \ and \ \ Dimitri Kusnezov\footnote{E--mail: {\tt
      dimitri@nst.physics.yale.edu}}}
\address{
  $^*$\it Dept. of
  Physics, Keio University, \\
  {\it 4---1---1} Hiyoshi, Kouhoku--ku,
  Yokohama 223--8521, Japan\\
  $^\dagger$\it Center for Theoretical Physics,
  Sloane Physics Lab, Yale University, \\
  New Haven, CT\ 06520-8120, U.S.A.
}
\maketitle
\begin{abstract}
  We study the dynamics of $\phi^4$ theory and the FPU $\beta$
  model under thermal gradients, from first principles.  We
  analyze quantitatively how local equilibrium and linear
  response are violated, paying special care to how we find
  observables that unambiguously display these violations.
  Relations between these quantities to equations of state are
  also examined.  Further, we discuss how we can approach
  similar dynamical problems in continuum quantum field theory.
  We analyze how close we are to obtaining the continuum
  results. 
\end{abstract}
\section{Introduction}
\label{sec:intro}
Non-equilibrium dynamics appears in almost all areas of physics,
from physics of the early universe, heavy ion collisions to
transport in matter.  In many of these situations, we would like
to find out how physical quantities behave in various quantum
field theories under non-equilibrium conditions.  In an ideal
world, we would like to analyze their dynamics from first
principles, with no approximations.  Let us first discuss what
this entails: We would use no dynamical assumptions, the
computations will be fully non-perturbative, transport
properties will not be restricted only to linear response and the 
non-equilibrium properties of the system arise dynamically if
some non-equilibrium conditions are imposed at the
boundaries. These are some of the requirements that need to be
satisfied. 

In practice, this is too ambitious. Usually, we resort to
computations of the Green--Kubo formulas for the transport
coefficients or use the Boltzmann equation, which is truncated
from the full BBGKY hierarchy.  The equations are solved
essentially in equilibrium so that the region of validity of the
results are unclear, not to mention that the results can not be
applied to situations far from equilibrium.  Even then, further
approximations need to be made.  For instance, the transport
coefficients of $\phi^4$ theory to {\it leading order} in the
coupling was completed only recently \cite{hosoya,jy}. The
reason for this is clear; the computations are quite involved
and results can not be obtained in closed form, except for its
asymptotic behaviors.

In this work, we study the non-equilibrium dynamics of two types
of theories, $\phi^4$ theory and the FPU $\beta$ model, in
$D=1,2,3$ spatial dimensions.  We shall work with  {\it
  classical} theories on the lattice and from {\it first
  principles}. In particular, we shall make {\it no } dynamical
assumptions and our computations will be fully non-perturbative.
While the $\phi^4$ theory has a bulk limit, the FPU $\beta$
model does not, and the temperature dependence of the transport
coefficients qualitatively differs.  It hardly needs to be
mentioned, but the $\phi^4$ theory is a classic prototypical
field theory and the FPU $\beta$ model is another classic model
which has been studied widely. These distinct theories in
various dimensions should allow us to gauge how applicable the
results are to more general theories. Putting these systems
under weak and strong thermal gradients, we shall study how we
can measure deviations from local equilibrium and analyze its
behavior quantitatively.  We also examine the validity of the
linear response law and analyze its violations quantitatively.
Further, we study the relations of these violations to the
thermodynamic properties of the systems, such as the equations
of state.  We then discuss how to extend these results to
quantum field theories.

The main reason we can analyze the systems from first principles
is because we are working with classical systems on a finite
lattice (of varying sizes).  In the classical theory, the
problem essentially reduces to computing the solutions to the
equations of motion for the degrees of freedom of the system. We
can perform this task without any approximations such as
perturbation theory by numerically integrating the equations of
motion.  While the behavior of a classical system is different
from that of a quantum system, the latter often reduces to the
former under certain conditions, such as for high
temperatures\cite{thermal-texts}.  Also, the classical theory
can play a major role in obtaining the results in the full
quantum theory, as can be seen from the results in the
electroweak theory at finite temperatures \cite{ew}.
Furthermore, we believe that one should understand the behavior
of classical systems first, before we can hope to understand the
full quantum theory. We work with lattice theories which we can
think of as a real lattice as in solids. We can also think of
the lattice as a discrete approximation of the continuum theory,
which is quite natural and corresponds to a theory with a
cutoff. Of course, it hardly needs to be mentioned that the
behavior of classical lattice systems is of interest on its own
right!
\section{The systems}
\label{sec:systems}
The Hamiltonians of the models we study are 
\begin{equation}
  \label{ham}
  H=\bfr\sum\left[ \half\bfr\pi^2+\half{(\nabla\phi\bfr)^2}+
  V(\bfr\phi)\right],\qquad
  V_{\phi^4}(\bfr\phi)={\bfr\phi^4\over4},\quad
  V_{FPU}(\bfr\phi)={(\nabla\phi\bfr)^4\over4}
\end{equation}
in $1$---$3$ spatial dimensions. The sum is taken over all
lattice points $\bf r$ and $\nabla \bfr\phi$ is the lattice
gradient.
As should be clear, the $\phi^4$ theory can be thought of as a
discretized version of the standard continuum $\phi^4$ theory.
The mass has been set to zero for simplicity here, although we
have also studied cases where the mass is non-zero.

In this work, we are interested in the steady state properties
of the systems under thermal gradients.  As such, we apply
thermostats at the boundaries $x=0,L$ and impose periodic
boundary conditions for the other directions for $D=2,3$.  It is
important to note that the thermostats are applied only at the
ends so that in the interior, $0<x<L$, the degrees of freedom
are {\it solely } those of the $\phi^4$ theory or the FPU model.
The thermostats we use are generalized variants of the
thermostats of Nos\'e and Hoover\cite{nh,bk}.  Explicit forms of
equations of motion including the thermostat degrees of freedom
can be found in \cite{ak-plb,ak-long}.  The thermostats are
deterministic, rather than stochastic, and the thermostats
effectively impose thermal distributions at the thermostatted
sites when time averaged.  A physical observable, $\cal O$, such
as correlation functions or currents, is computed by sampling
them along the classical trajectory of the whole system in the
phase space and taking its time average, $\ave{\cal O}$.  We
studied lattice sizes in the $x$ direction up to $L=10000$ for
$D=1$ and up to $L=2000$ and transverse lattice sizes of
$L_\perp=3\sim20$ when $D=2,3$.

Since we will be interested in thermal transport, we shall need
an expression for the energy flow, 
\begin{equation}
  \label{t01}
  \c T^{0x}_{\phi^4}=
  -\bfr\pi\nabla_x\bfr\phi,\qquad
  \c T^{0x}_{FPU}=
    -\left(\bfr\pi\nabla_x\bfr\phi
      \right)\left[1+(\nabla\phi\bfr)^2\right]
\end{equation}
which satisfies the usual continuity equation.
\section{Near equilibrium behavior}
\label{sec:neareq}
When the two end point temperatures, are equal, $T_1=T_2$, the
interior of the system thermalizes at the same temperature.  This
can be explicitly confirmed by checking that the momentum
distribution is Maxwellian.  The local temperature is defined by
$T({\bf r})=\ave{\pi_{\bf r}^2}$, which is called the ideal gas
temperature.  This is well defined and is a robust local
measure, since the Hamiltonians in \eqnn{ham} are quadratic in
$\pi_{\bf r}$ and $\pi_{\bf r}$ interact with the neighboring
sites only indirectly through $\phi_{\bf r}$.

\begin{figure}[htbp]
  \begin{center}
    \leavevmode
    \epsfysize=5.3cm\epsfbox{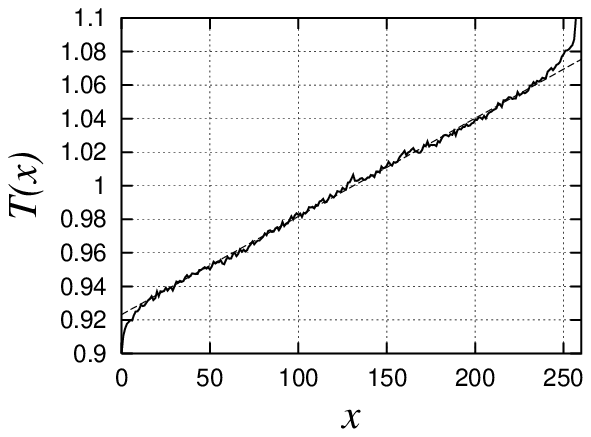}\hspace{1cm}
    \epsfysize=5.3cm\epsfbox{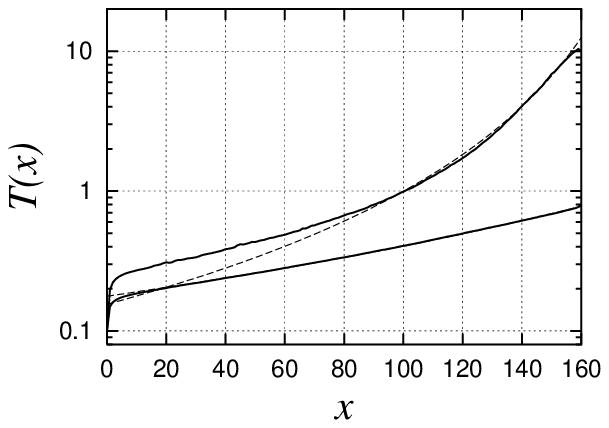}
    \caption{(Left) A linear temperature profile (solid) and its
      linear fit (dashes) in the FPU model for
      $L=260$, $(T_1,T_2)=(0.9,1.1)$, $D=1$.  (Right) Curved
      profiles (solid) in $\phi^4$ theory with the linear response
      predictions (dashes) for $L=160$,
      $(T_1,T_2)=(0.1,0.8),(0.1,10)$, $D=1$.  We see that the
      linear response prediction is applicable for the smaller
      gradient case, but not for the other.}
    \label{fig:profs}
  \end{center}
\end{figure}
\begin{figure}[htbp]
  \begin{center}
    \leavevmode
    \epsfysize=5.3cm\epsfbox{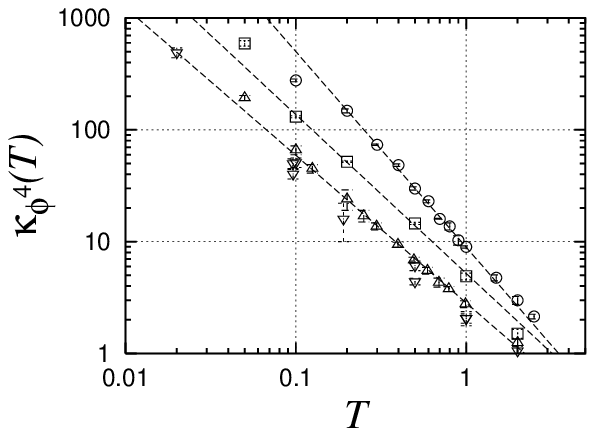}\hspace{1cm}
    \epsfysize=5.3cm\epsfbox{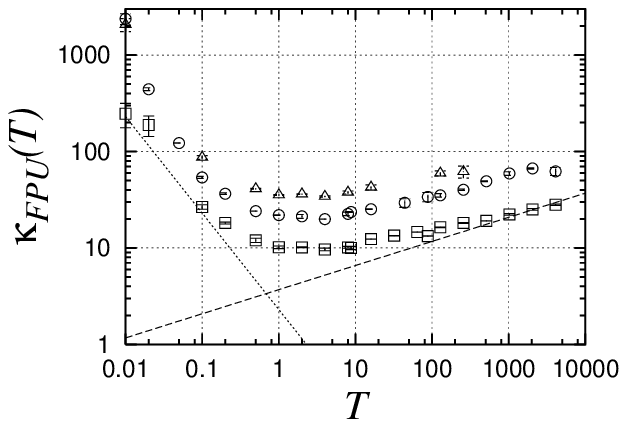}
    \caption{(Left) Temperature dependence of $\kappa$ for the $\phi^4$
      theory in $D=1\ (\bigtriangleup)$ $D=2\ (\Box)$, and $D=3\ 
      (\circ)$ with the power laws of \eqnn{tc-phi} in
      dashes. The Green--Kubo results for $D=1$ are also shown
      ($\bigtriangledown$).  (Right) Behavior of $\kappa$ for
      the $D=1$ FPU model for $L=16\ (\Box)$, $L=128\ (\circ)$
      and $L=512\ (\bigtriangleup)$.  The asymptotic behaviors of
      $1/T$ (dots) and $T^{1/4}$ (dashes) are also
      plotted. The $\phi^4$ theory has a bulk limit, whereas the
      FPU model does not.}
    \label{fig:tc}
  \end{center}
\end{figure}

When we make the two boundary temperatures slightly different,
we recover a straight temperature profile which can be well
described by the Fourier's law,
\begin{equation}
  \label{fourier}
  \ave{{\c T}^{0x}}=-\kappa(T)\nabla_x T
\end{equation}
as can be seen in \figno{profs} (left).\footnote{While it will
  not be discussed here, in general, temperature jumps will
  arise at the boundaries which can be understood\cite{ak-pla}.}
By choosing various small temperature differences around the
same central temperature, we directly extract the thermal
conductivity $\kappa$ for a given temperature in each system at
each lattice size from measurements of the energy flow and the
temperature gradient, using \eqnn{fourier}.  In the $\phi^4$
theory, we find a bulk limit exists for $\kappa$, namely the
limit $\kappa\ (L\rightarrow\infty)$
exists\cite{ak-plb,ak-long}.  In the FPU model, the bulk limit
does {\it not } exist\cite{fpu-later,ak-fpu}. The results can be
summarized for the $\phi^4$ theory as (see \figno{tc}~(left))
\begin{equation}
  \label{tc-phi}
  \kappa(T)={A\over T^\gamma},\hspace{3cm}
  \matrix{ & A & \gamma\cr
    D=1& 2.8(1) & 1.32(2)\cr
    D=2&5.2(4)&1.42(6)\cr
    D=3&8.9(5)&1.75(5)\cr}
\end{equation}
For the FPU model, the behavior is more complicated as can be
seen in \figno{tc}~(right); $\kappa$ depends on L and $T$
dependence is no longer a simple power.  The $L$ dependence is
compatible with a simple power law and the asymptotic behavior
\begin{equation}
  \label{tc-fpu} 
  \kappa \simeq\cases{  1.2 L^\delta T^{-1}&   $(T\alt 0.1)$\cr
    2 L^\delta T^{1/4}&$ (T\agt 50)$\cr},\qquad
  \delta=0.37(3).
\end{equation}
can be understood from scaling arguments\cite{ak-fpu}.

We can also compute the thermal conductivity {\it independently}
of the Fourier's law~\eqnn{fourier}
through the integral of the autocorrelation of the energy
current using the Green--Kubo formula. The results agree with the
direct computation from Fourier's law.  For $D=1$ $\phi^4$
theory, the Green--Kubo results are shown in \figno{tc}~(left)
along with those from direct computations\cite{ak-pla}.  We note in passing
that the applicability of the Green--Kubo formalism is regarded
as being most subtle for lower dimensional systems due to the
existence of persistent correlations\cite{tails}.

As we make the temperature gradients larger, the temperature
profiles become visibly non-linear, as in
\figno{profs}~(right). This does { not} necessarily
signal the breakdown of linear response.\footnote{By ``linear
  response'', we refer strictly to the validity of
  \eqnn{fourier}. In particular, we do not presume
  that the existence of a bulk limit.}%
Since $\kappa$ depends on the temperature ({\it cf.}
\figno{tc}), as the temperature changes, so does the gradient,
and it can do so within the same system\cite{profs}. This
reasoning leads to the following formula for the temperature
profile by applying the linear response law~\eqnn{fourier} {\it
  locally}\cite{ak-plb}.
\begin{equation}
  \label{tprof}
  T(x)=T_1\left[1-{x\over L}+ \left(T_2\over T_1
    \right)^{1-\gamma}
      {x\over L}\right]^{1\over1-\gamma}  
\end{equation}
Here, we assumed that the temperature dependence of the thermal
conductivity can be described by $\kappa(T)=AT^{-\gamma}$, which
is true for the $\phi^4$ theory and for most temperature regions in
the FPU model.  It should be noted that the profile has the
scaling behavior that it depends on $x$ only through $x/L$.
This formula for the profile works quite well when the we are
not too far from equilibrium. For example in \figno{profs}
(right), the theoretical curve~\eqnn{tprof} along with the
profile is presented and we see that it provides an excellent
description of the curved temperature profile for not so large
gradients. We shall turn to the question of quantifying how far
we are from equilibrium and how good the description is, in the
next section.
\section{Breakdown of local equilibrium and linear response}
\label{sec:breakdown}
When the system moves further away from equilibrium, the
formula~\eqnn{tprof} is no longer adequate to describe the
temperature profile, as in \figno{profs}~(right).  One reaction
might be to employ some kind of `non-linear' response theory.
However, we would first like to understand the overall behavior
of the system quantitatively and judge whether such a
description is valid.  To this end, we would like to also assess
if local equilibrium is satisfied so that for instance, the
temperature is well defined.  For this, we need an explicit
measure of how well the local equilibrium holds, which we find
in the cumulants of $\bfr\pi$. Since the Hamiltonian of the
system~\eqnn{ham} is quadratic in $\bfr\pi$, the thermal
distribution should be Maxwellian {\it if} local equilibrium
holds.  The cumulants provide an unambiguous local measure of
local equilibrium breakdown.  We shall also need a local measure
of how ``far'' we are from equilibrium and for this, we use
$\nabla_xT/T$. We note that it is natural to use the rescaled
gradient here. This is quite evident in \figno{profs}~(right);
the linear response prediction is less applicable in the small
$T$ region, wherein $\nabla_x T$ is smaller but $\nabla_x T/T$
is larger.

Before proceeding further, we would like to discuss the general
picture: We find that a physical observable, $\c A$, generically
deviates from its local equilibrium value in a manner
\begin{equation}
  \label{o-le}
  \delta_{\cal A}\equiv
  {\delta {\cal A}\over{\cal A}}=
  C_{\cal A}\left(\nabla_x T\over T\right)^2
  +C'_{\cal A}\left(\nabla_x T\over T\right)^4
  +\ldots
\end{equation}
While this seems quite natural since the intrinsic behavior
should not depend on the direction of the gradient, it has not
been established analytically. Furthermore, the behavior of the
coefficients is more subtle than what one would expect, as we
shall see below.

\begin{figure}[htbp]
  \begin{center}
    \leavevmode
    \epsfysize=5.3cm\epsfbox{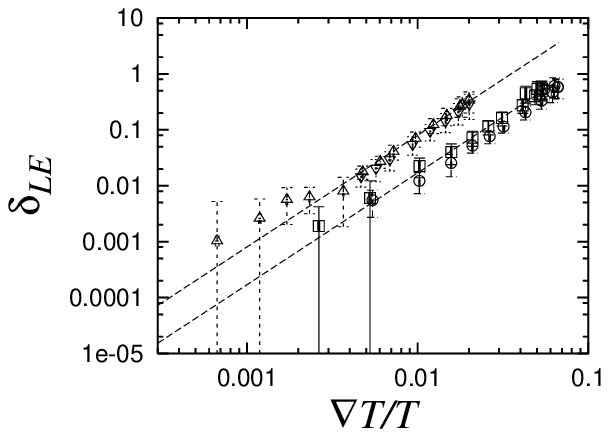}\hspace{1cm}
    \epsfysize=5.3cm\epsfbox{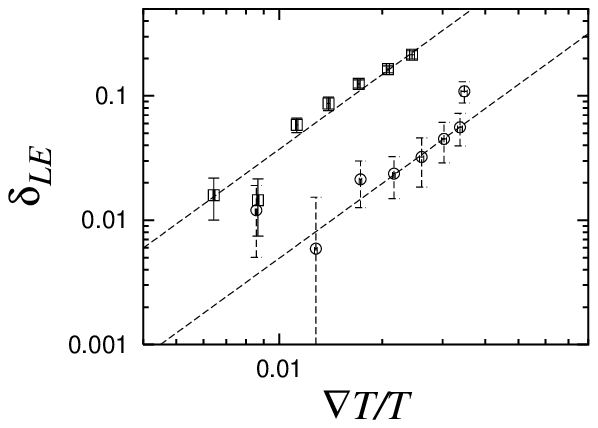}
    \vspace{0.2cm}
    \caption{Violations of local equilibrium and linear
      response. (Left) $\delta_{LE}$  for FPU model in $D=1$
      with $L=16,\ T=8.8\ (\Box),\ T=88\ (\circ)$,  $L=64,\
      T=8.8\ (\bigtriangleup),\ T=88\ (\bigtriangledown)$
      together with the quadratic behavior (dashes). $L$
      dependence is obvious but $T$ dependence is weak.
      (Right) $\delta_{LR}$ in $\phi^4$ theory, $D=1$ at $T=1$
      for $L=40\ (\circ)$ and $L=80\ (\Box)$ along with the
      quadratic behavior (dashes). $L$ dependent
      behavior is evident.}
    \label{fig:violations}
  \end{center}
\end{figure}

Let us analyze first the violations of local equilibrium and 
linear response. For deviations from local equilibrium, we use
the observable, $\cum{\pi^4}/T^2$ where the fourth cumulant is
$\cum{\pi^4}\equiv\left\langle\pi^4\right\rangle -
3\left\langle\pi^2\right\rangle^2 $.  Likewise, for linear
response, we use the observable $(\c T^{0x}-\kappa\nabla_x
T)/\kappa\nabla_x T$. We show some of the results in
\figno{violations}.  These computations need to be performed
with care since we are looking at differences.

{}From these analyses, we extract the coefficients $C_{LE},C_{LR}$
in \eqnn{o-le} for local equilibrium and linear response,
respectively.  $C_{LE},C_{LR}$ can in general depend on $T$ and
$L$. We have computed these coefficients at various $T$ and $L$
for both the $\phi^4$ theory and the FPU model in various
dimensions. While we do not have yet a global picture of how
these coefficients behave, a few comments are in order.  For the
$\phi^4$ theory, which has a bulk limit, we would naively expect
that these coefficients will be independent of $L$. However,
this turns out to be {\it not} the case.  For instance, at
$T=1$, $C_{LE}=aL$ with $a=3,2,4$ for $D=1,2,3$.  There is thus
a strong dependence on the size of the system, so that a simple
local understanding of \eqnn{o-le} does not seem possible. In
general, both for the $\phi^4$ theory and for the FPU model in
$D=1,2,3$ and for various $L$, we find that
$\delta_{LE}\sim\delta_{LR}$.  This means, in particular, that a
simple non-linear response theory is not applicable to these
systems since the breakdown of local equilibrium also needs to
be considered simultaneously. We note that a priori, it could
have been such that $\delta_{LE}$ was much smaller that
$\delta_{LR}$, in which case a non-linear response theory would
have seemed quite appropriate. Generically, we find the
dependence of $C_{LE},C_{LR}$ on $L$ to be close to linear, and the
temperature dependence to be weak, which supports the rescaled
gradient $\nabla_x T/T$ as a natural measure of how far we are
from equilibrium.  We have further computed the deviations in
the equations of state for pressure and local energy density
from local equilibrium relations.  We find that the deviations
can again be understood using the description in \eqnn{o-le}.
\section{Towards quantum field theory in non-equilibrium}
\label{sec:qft}
As we have seen in the previous sections, in  classical
lattice systems, we can ask practically any question regarding
the physics behavior of non-equilibrium systems and obtain an
answer, at least under steady state conditions.  We are not
restricted to weak coupling, nor do we have to assume linear
response theory and we can be arbitrarily far from equilibrium.
Clearly, we want to extend this situation to quantum field
theories.
Ideally, we would like to solve quantum dynamics in a similar
fashion, leading to the solution of quantum field theory in
non-equilibrium, in the continuum limit. This is, of course, too
ambitious at this point; even for moderately large systems (say
$\sim100$ degrees of freedom), it is computationally prohibitive
to carry out such a program. Our goal here will be more modest;
we would first like to apply the results of the classical theory
to quantum field theory and find out how they fit in. We shall
see that even this task is quite non-trivial.  Typically,
classical results are applicable to quantum theories in some
region or in some limit \cite{thermal-texts,ew}. At first sight,
the prospect seems quite promising for the lattice $\phi^4$
theory; amongst other features, the theory has a bulk limit with
the behavior appearing already for system sizes of order hundred
or less for moderate temperatures and the temperature profile
\eqnn{tprof} has a smooth continuum limit.

However, in practice, we immediately encounter an obvious
problem: when one tries to take the naive continuum limit by
taking the lattice spacing to zero, quantities such as the
energy density, thermal conductivity diverge.  They diverge for
the trivial reason that the system will have an infinite number
of oscillators in the unit volume.This is the
``Rayleigh--Jeans'' problem, known from the early days of the
20th century which afflicts all classical theories in the
continuum. While it might seem that this will make it impossible
to obtain finite results in any classical continuum theory at
finite temperatures, it has been demonstrated that one can
obtain finite results for physical observables in the classical
continuum theory that matches with the quantum results in the
appropriate region\cite{ew,aarts}.  This is done by renormalizing
the theory classically using the appropriate matching
conditions.

\begin{figure}[htbp]
  \begin{center}
    \leavevmode
    \epsfysize=5.3cm\epsfbox{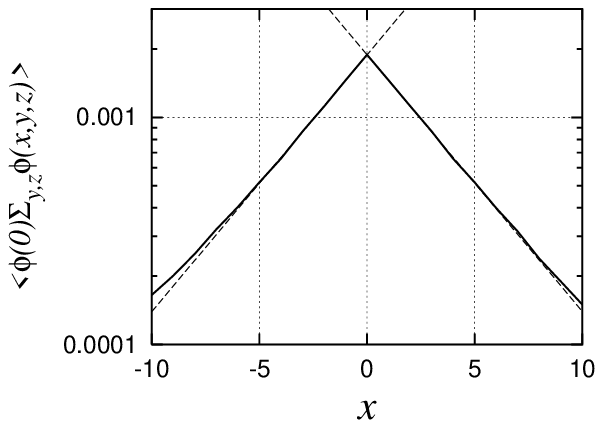}\hspace{1cm}
    \epsfysize=5.3cm\epsfbox{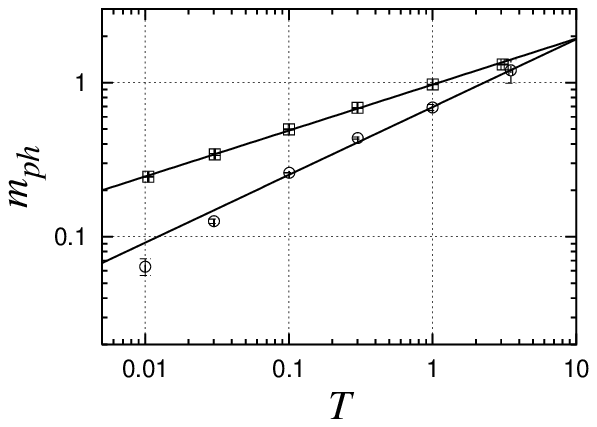}
    \vspace{0.2cm}
    \caption{(Left) The behavior of the correlation function in
      \eqnn{physmass} (solid) and its exponential behavior
      (dashes) for the $\phi^4$ theory.  The lattice size is
      $23\times10\times10$ at $T=0.1$ in this example. (Right)
      The behavior of $m_{ph}$ with respect to $T$ for $D=1\ 
      (\Box)$,$D=3\ (\circ)$. The power behavior (solid)
      describes the results quite well.}
    \label{fig:physmass}
  \end{center}
\end{figure}

Next point to be resolved before we can start applying our
results to quantum field theory is the continuum limit. How far
are we from the continuum limit and how do we know?  While we do
not have a definitive answer to this difficult question, we can
study the physical mass in lattice units, $m_{ph}a$, which
effectively measures how big our lattice mesh is in physical
units.  (The lattice size, $a$, has been set to one here.)  To
avoid lattice artifacts, we would like this quantity to be
small. The physical mass can be measured from the correlation
function as
\begin{equation}
  \label{physmass}
  \Big\langle\phi(0)\,\sum_{{\bf r}_\perp}\phi(x,{\bf
  r}_\perp)\Big\rangle \sim e^{-m_{ph}|x|}
\end{equation}
where $\bf r_\perp$ denotes the coordinates transverse to
$x$. In \figno{physmass} (left), we study the correlation
function averaged in the transverse direction. We see that the
correlation function indeed does decay exponentially, which
allows us to extract the physical mass.  We find, as in
\figno{physmass} (right), that the physical mass has a
temperature dependence which we find to be
\begin{equation}
  \label{physmass-behavior}
  m_{ph}= 
\cases{ 0.97(2) T^{0.30(1)}& $D=1$ \cr
      0.69(5) T^{0.44(3)}& $D=3$ \cr}
\end{equation}
To put things in perspective, in lattice QCD, the state of the
art simulations are typically run with $m_{ph}a=0.1\sim0.5$. So,
for $T\lesssim1$, we should be able to overcome lattice
artifacts.

One more aspect needs to be addressed: In the continuum limit,
when the lattice no longer exists and the theory becomes
relativistic, we have to decide relative to what reference frame
we have transport in the bulk.  Here, a natural situation is
provided if we consider a system with more conserved charges.
For instance, it is quite straightforward to extend our analysis
to a complex scalar field theory with U$(1)$ symmetry. In this
case, we can address energy transport in the system with no net
charge flow, which would allow us to extract the thermal
conductivity in the usual sense. Since our methods can be used
in practically any classical lattice system, we should be able
to address the problem of relativistic transport, with 
judicious choice of models.

While more work is definitely needed to apply our results to the 
appropriate regimes of quantum field theory, none of the
obstacles seems insurmountable.  The lattice artifacts should
be reasonably small for $T\lesssim1$, the divergences of the
classical continuum theory can be overcome and the relativistic
limit is quite compatible.
\section{Discussions}
\label{sec:disc}
In summary, we have been able to obtain the physical behavior of
$\phi^4$ theory and the FPU $\beta$ model under thermal
gradients, near and far from equilibrium. Close to equilibrium,
we see that linear response theory works well and local
equilibrium description is applicable.  As we move away from
equilibrium, linear response breaks down, but so does local
equilibrium.  We have measured {\it quantitatively} how much
they deviate from equilibrium values and they are rather similar
in all the systems we studied. The deviation of these quantities
from the equilibrium behavior is quadratic in the thermal
gradient, as in \eqnn{o-le}.  This behavior is rather generic in
the systems we studied, but no analytic explanation of this
behavior is known. It is interesting to note that in the
non-linear dependence of the viscosity on the shear rate, it was
initially argued analytically and further found numerically that
the behavior is {\it not } quadratic but rather has a
non-analytic behavior with a power of $3/2$\cite{shear1}.
Research in this area is still in progress and the question
whether the dependence is analytic still seems quite open
\cite{shear2}.

We have further discussed how to apply our results to quantum
field theory. Obviously, our results do not incorporate
essentially quantum behavior so that they can be applied only to
certain regimes in the quantum theory. Even though the task is
non-trivial, we found that the program is quite feasible.

Much more needs to be done: While we can answer practically any
question thrown at us regarding the steady state behavior of
classical lattice systems under thermal gradients, the results
by themselves lack analytic understanding. We have tried to fill
in the missing links by analyzing these results, but we feel
that a deeper understanding of the phenomena and their relation
to the intrinsic non-equilibrium dynamics is highly
desirable. For instance, as we mentioned above, even the
quadratic behavior of the deviations from equilibrium is
something that has not been established analytically in any
model.  While this behavior might be intuitive and
understandable, the fact that we have the breakdown of local
behavior in these deviations is quite subtle and needs to be
explained.  In applying the classical results to the quantum
theory, we still need to complete the program. Furthermore, if
one wanted to perform even moderately large scale {\it quantum}
simulations from first principles, how to approach it is still
unclear.  Also, even in classical systems, we have not dealt
with transient phenomena, which is an interesting and important
avenue of study.  We analyzed the steady state physics as a
first step, but many non-equilibrium phenomena, such as the
theory of the early universe or heavy ion collisions, are
intrinsically time dependent.  We feel that there is still much
more interesting physics to be uncovered in the area of
non-equilibrium systems, even classically.
\par\noindent{\bf Acknowledgements: }We would like to thank
Gert Aarts, Jan Smit and Larry Yaffe for stimulating discussions
regarding the material in the last section.

\end{document}